\documentclass[pra,aps,twocolumn,superscriptaddress,showpacs,showkeys,keywords]{revtex4-1}
\usepackage{amsmath,amsthm,amssymb,graphicx,xcolor}
\usepackage{graphicx}

\usepackage[unicode=true]{hyperref}
\hypersetup{
	colorlinks=true,       		
	linkcolor=blue,          	
	citecolor=red,            
	urlcolor=magenta,           	
}

\newtheorem*{theorem*}{Theorem}

\newtheorem*{corollary*}{Corollary}

\newtheorem*{lemma*}{Lemma}

\newtheorem*{proposition*}{Proposition}
\theoremstyle{definition}

\newtheorem*{definition*}{Definition}
\theoremstyle{remark}

\newtheorem*{remark*}{Remark}

\begin{document}
	
\title{Wave and particle properties can be spatially separated in a quantum entity}
	\author{Pratyusha Chowdhury}
	\email{pchowdhury@iitg.ac.in}
	\affiliation{Indian Institute of Technology Guwahati, Guwahati 781039, Assam, India}

	\author{Arun Kumar Pati}
	\email{akpati@hri.res.in}
	\affiliation{Quantum Information and Computation Group, Harish-Chandra Research Institute, Chhatnag Road, Jhunsi, Allahabad 211 019, India}
	
	\author{Jing-Ling Chen}
	\email{chenjl@nankai.edu.cn}
	\affiliation{Theoretical Physics Division, Chern Institute of Mathematics, Nankai University, Tianjin 300071, People's Republic of China}
	
	\date{\today}
	\begin{abstract}
Wave and particle are two fundamental properties of \emph{Nature}. The wave-particle duality has indicated that a quantum object may exhibit the behaviors of both wave and particle, depending upon the circumstances of the experiment. The major significance of wave-particle duality has led to a fundamental equation in quantum mechanics, the Schr{\" o}dinger equation. At present, the principle of wave-particle duality has been deeply rooted in people's hearts. This gives rise to a common sense perception that wave property and particle property coexist simultaneously in a quantum entity, and these two physical attributes cannot be completely separated from each other. In classical physics, a similar common sense is that a physical system is inseparable from its physical properties. However, this has been recently challenged and beaten by a quantum phenomenon called the ``quantum Chesire cat", for which a cat and its grin can be separated spatially. In this work, we propose a thought experiment based on the similar technology of quantum Chesire cat. We find that wave and particle attributes of a quantum entity can be completely separated, thus successfully dismantling the wave-particle duality for a quantum entity. Our result is still consistent with the complementarity principle and deepens the understanding of quantum foundations.
	\end{abstract}
	
	
	\maketitle

\section{Introduction}

Whether light is a wave or a particle has been a long-term debate, which can be traced back to Newton's corpuscular theory and Huygens's wave theory
	in the 17th century~\cite{100Years,Feynman1965}. The phenomena of interference, diffraction and polarization have convinced people that light could be fully described by
	a wave, but the appearance of photoelectric effect has introduced indisputable evidence that light exhibited particle property in the microscopic world ~\cite{Einstein1905}.
	As a compromising result, the wave-particle duality of light was eventually and widely accepted~\cite{Wheeler1984}. In 1923, the French physicist Louis de Broglie generalized
	the viewpoint of wave-particle duality from light to electron, and also to all other matters~\cite{deBroglie1,deBroglie2}. He boldly proposed that electrons with
	momentum $p$ should exhibit the wave properties with an associated wavelength $\lambda= h/p$, with $h$ being Planck's constant. Later on, Davisson and Germer
experimentally confirmed de Broglie's hypothesis about the wave-particle duality of matters by observing the electron diffraction effects~\cite{Davisson1927}. Subsequently, the wave-particle duality has laid the foundation stone for the development of new quantum theory, e.g., it has stimulated the establishment of Schr{\" o}dinger's equation, a fundamental equation in quantum mechanics.
	
	Even after so many years of the development of quantum mechanics, the wave-particle duality is still one of the most intriguing features of the theory. Such a duality
	supposes that a quantum particle is accompanied by a wave, i.e., both the particle and the wave are assumed to exist objectively. The duality has its own roots in the
	complementarity principle \cite{bohr}. It has been studied extensively in the past, and still it does not stop to amaze us with its profound implications.
	The most dramatic consequence of the wave-particle duality is the quantum interference which
	is displayed on a screen when we send photons or particles in a double slit set-up. The remarkable thing is that this quantum interference occurs even if only one
	particle is sent at a time and the particle seems somehow to pass through both slits at once, thus leading to interference. How each particle passes through both the slits
	is still a mystery. It may be noted that to explain
	the quantum interference, it has been postulated that when the quantum entity impinges on the beam splitter the particle may be going along one path but the wave is
	divided and travels along both the paths. The wave that goes along the arm where the particle is not present is called an `empty wave' \cite{selleri}.
	Although there have been long drawn debates on empty waves, i.e.,  the waves that do not contain the associated particle properties, yet this proposal seems to
	confuse many people, and is accepted by some and disregarded by others (see for example \cite{selleri,hardy}).
	It may be the case that the nature of quantum entity may be different than what
	the wave-particle duality has actually depicted us \cite{jan}. For example, the wave and the particle we associate with a quantum entity is not same as the wave and the
	particle
	that we see in the classical world. Recently, there has been an attempt to quantify the nature of particle using resource theoretic framework \cite{indra} where it was proposed
	that for each quantum entity there are myriads of waves and particles.

	Another intriguing aspect of quantum mechanics is the concept of weak measurement \cite{aharonov1988result,aharonov1990properties,jozsa2007complex,
	hosten2008observation,lundeen2011direct} with suitable pre- and post-selections. Using the weak measurement formalism, it has been suggested that the
	\emph{quantum Cheshire cat}~\cite{Aharonov2013} can be a possibility where a cat and its grin can be separated spatially. In quantum mechanics, this
	essentially means that with suitable pre-and post-selected states one can spatially separate the spin of a particle and the particle itself. In recent years,
	this work has raised lot of questions about separating an attribute of a physical system from the system itself -- a concept that seemed only possible in fictions~\cite{cheshire1865}. However,
	when this becomes a scientific result, then it is bound to cater attention from scientists all over the world. Over last few years, lot of works have been done in this
	area to unravel the mysteries of the nature~\cite{Ibnouhsein2012,Guryanova2012,
Lorenzo2012,pan,eli,denkmayr2014observation,ashby2016observation,
liu2020experimental,atherton2015observation,das2019can,das2019teleporting,sen,correa2015quantum,atherton2015observation,bancal2014quantum,duprey2018quantum}. It should be
further
noted that this phenomenon has not only become a theoretical construct, but also been experimentally verified~\cite{denkmayr2014observation,ashby2016observation,
liu2020experimental,atherton2015observation}.
	
	The enduring view about the wave-particle duality has suggested that a quantum entity behaves like both a wave and a particle. Suppose one can spatially separate the wave
	property and the particle property of lights or electrons, this immediately gives rise to some fundamental questions: Can one still observe the interference fringes on
	the screen when he adopts the lights with solely particle property to perform the Young-type double-slit experiments? Can one still observe the photoelectric effect when
	he adopts the lights with only wave property? Can one still observe the diffraction effects when he adopts the electrons with solely particle property to perform the
	corresponding experiments?, and so on. Undoubtedly, to answer the above questions, a crucial step is developing a technology to completely separate the wave property and the particle property for a single physical entity.
	
	In this work, we intend to investigate whether any profound implication can be drawn by linking the wave-particle duality and the quantum Cheshire cat. We shall propose a thought experiment with the help of quantum Cheshire cat, such that it is possible to spatially separate the particle aspects from the wave aspects for 	a quantum entity by using the suitable pre- and post-selections. We will show that the particle attribute is not displayed in one arm of the interferometer and the wave attribute is not displayed in another arm of the interferometer. Nevertheless, we will show that the the quantum entity respects a new complementarity. Conclusion and discussions will be made in the end.
	
	\begin{figure}[t]
		\centering
		\includegraphics[width=80mm]{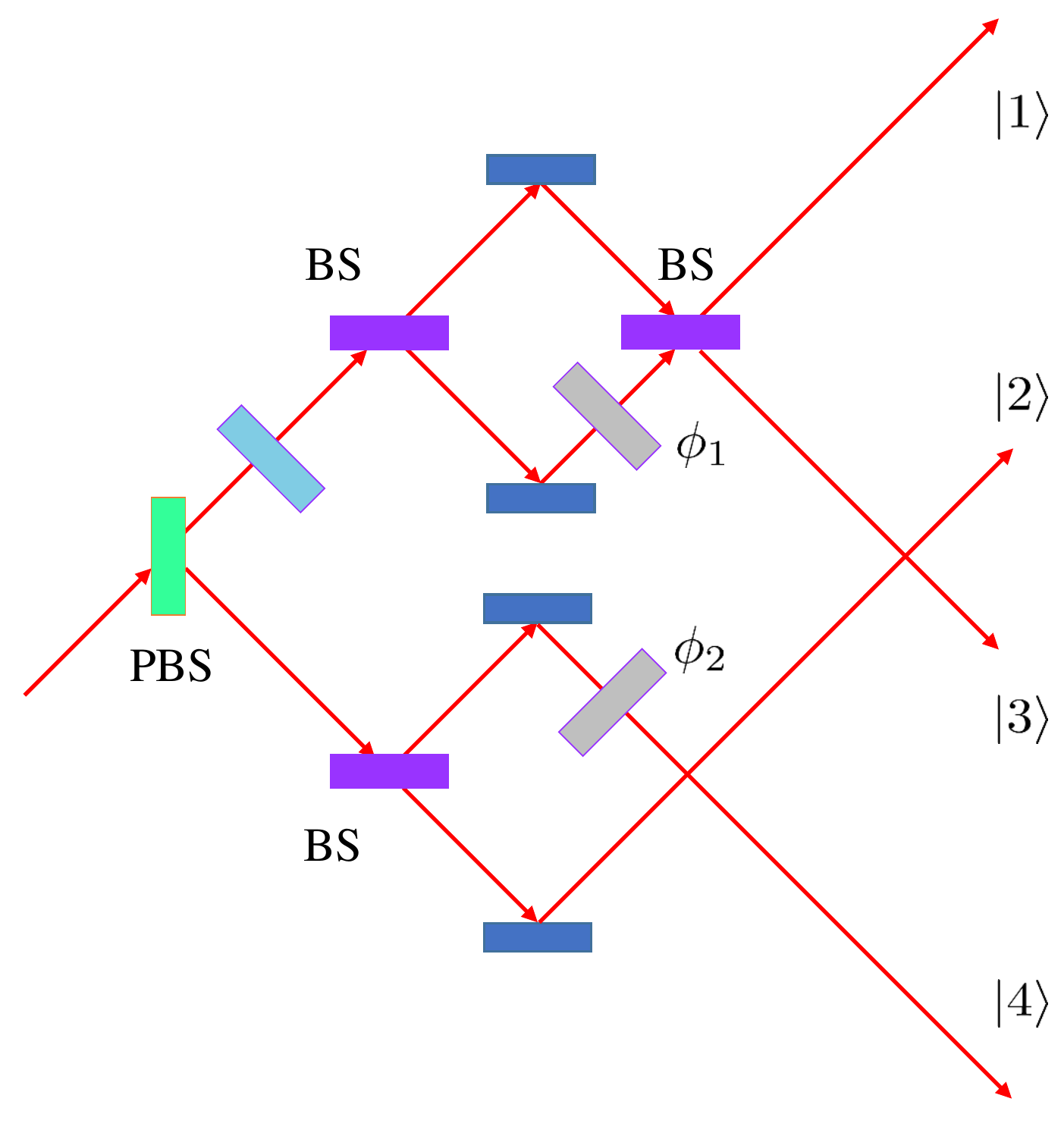}
		\caption{Schematic illustration of the wave-particle toolbox.}\label{fig1}
	\end{figure}

\section{Theoretical Framework}
In a recent paper~\cite{rab2017entanglement}, an outstanding progress has been made by presenting an experimental setup called the wave-particle (WP) toolbox. A schematic illustration of the toolbox can be found in Fig.\,\ref{fig1}. The conversion from the coherence superposition of the polarization states to the coherence superposition of the wave and particle entities exploits the wave-particle box. The mode conversion happening on the PBS reads $|\psi_{\rm in}\rangle = (\cos\alpha|H\rangle + \sin\alpha|V\rangle)|a\rangle 	\xrightarrow{\rm PBS} |\psi_1\rangle=\cos\alpha|H\rangle|1\rangle + \sin\alpha|V\rangle|2\rangle$, 	where ${\rm PBS}$ denotes the polarizing beam-splitter, the parameter $\alpha$ can be adjusted by a half-wave plate, and $|{H}\rangle$ and $|V\rangle$ denotes the horizontal and vertical polarization states, respectively. Now a half-wave plate (HWP) acts on the first path, hence $|H\rangle \rightarrow |V\rangle$, then the state $|\psi_1\rangle$ becomes the state $|\psi_2\rangle=$ $ |V\rangle(\cos\alpha|1\rangle + \sin\alpha|2\rangle)$,
	where $|n\rangle$ represents a state of a photon traveling along the $n$-th path. We can ignore the polarization degree of freedom in $|\psi_2\rangle$. Then, each path further bifurcates at a balanced beam splitter (BS). Hence
	$|\psi_2\rangle$ $\xrightarrow{ BS }|\psi_3\rangle$, where $|\psi_3\rangle=\cos\alpha[\frac{1}{\sqrt{2}}(|1\rangle + e^{i\phi_1}|3\rangle)]
	+ \sin\alpha[\frac{1}{\sqrt{2}}(|2\rangle+e^{i\phi_2} |4\rangle)]$,	and $\phi_1, \phi_2$ are the relative phases introduced by the phase shifters placed in paths 3 and 4. Now paths 1 and 3 are recombined by a beam splitter again. The phase shift between paths 1 and 3 is chosen to induce the following mode conversion: $|1\rangle \xrightarrow{BS} \frac{1}{\sqrt{2}}(|1\rangle+|3\rangle)$, $|3\rangle \xrightarrow{BS} \frac{1}{\sqrt{2}}(|1\rangle-|3\rangle)$. This leads to $\frac{1}{\sqrt{2}}(|1\rangle + e^{i\phi_1}|3\rangle) \xrightarrow{BS}  e^\frac{i\phi_1}{2}(\cos\frac{\phi_1}{2}|1\rangle-i\sin\frac{\phi_1}{2}|3\rangle) = |W\rangle$.
Then one has $|\psi_3\rangle \rightarrow |\psi_{\rm out}\rangle=\cos\alpha|W\rangle+ \sin\alpha|P\rangle$, with $|P\rangle  = \frac{1}{\sqrt{2}}(|2\rangle+e^{i\phi_2}|4\rangle)$.

Thus, due to such a toolbox, for a single photon prepared initially in a polarization state $|\psi_{\rm in}\rangle = \cos\alpha|H\rangle + \sin\alpha|V\rangle$, one obtains finally the output state as $|\psi_{\rm out} \rangle = \cos\alpha|W\rangle + \sin\alpha|P\rangle$, which is
a wave-particle superposition state of a single photon. Here, $|W \rangle$ and $|P\rangle$ denote respectively the wave and particle states as follows:
	\begin{eqnarray}\label{WP1}
&&	|W\rangle \equiv  |{\rm{Wave} }\rangle  = e^\frac{i\phi_1}{2}(\cos\frac{\phi_1}{2}|1\rangle- i\sin\frac{\phi_1}{2}|3\rangle),\nonumber\\
&& |P\rangle \equiv |{\rm{Particle}} \rangle = \frac{1}{\sqrt{2}}(|2\rangle+e^{i\phi_2}|4\rangle).
	\end{eqnarray}
Operationally these states represent the capability ($|W\rangle$) and incapability ($|P\rangle$) of the photon to produce interference. Here, $|n\rangle, n\in\{1, ..., 4\}$, denotes the $n$-th output mode from the wave-particle toolbox, and $\phi_1$ and $\phi_2$ are two controllable phase shifts in the toolbox.
If we represent the state of the photon as $|W\rangle$, then the probability of detecting the photon in the path $(n = 1, 3)$ depends on the phase $\phi_1$. In this case, the photon must have traveled along both paths simultaneously, thus revealing its wave behavior. If we represent the state of the photon as $|P \rangle$, then  the probability to detect the photon in the path $(n = 2, 4)$ is $\frac{1}{2}$ and does not depend on the phase $\phi_2$. In this case, the photon must have travelled only one of the two paths showing its particle behavior. In our setup, for simplicity, the phase shifters in the paths are the same, i.e.,  $\phi_1 = \phi_2$.

\begin{figure}[t]
		\includegraphics[width=80mm]{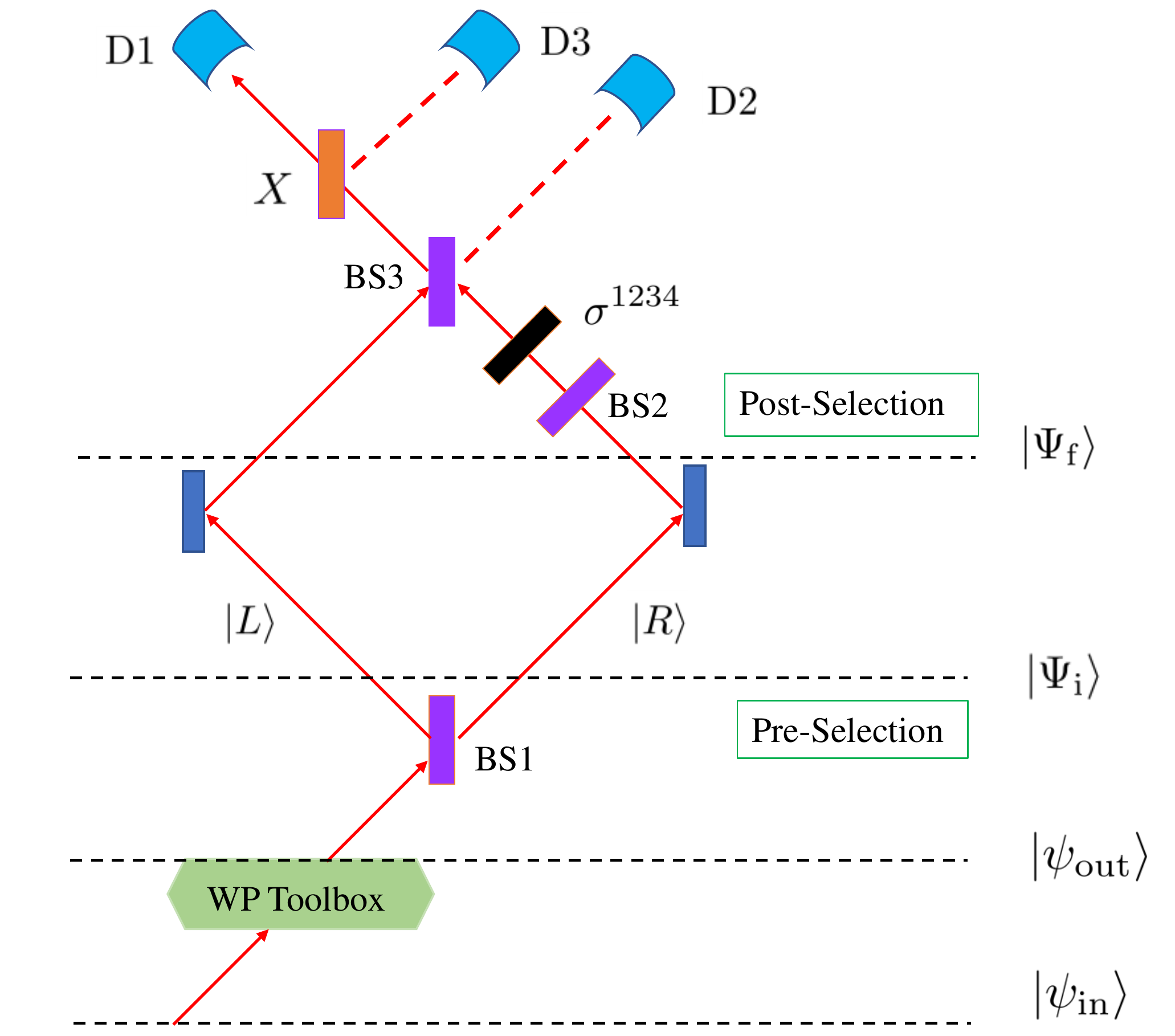}\\
		\caption{Illustration of spatially separating the wave property and the particle property of a single photon.
		}\label{fig2}
	\end{figure}

Following Ref. \cite{rab2017entanglement}, the illustration of spatially separating the wave and particle properties of a single photon is given in Fig.~\ref{fig2}. To separate the wave and particle properties, firstly we need to choose the pre-selected state as
	\begin{eqnarray}\label{pre-selestate}
	|\Psi_{\rm i} \rangle &=&\frac{1}{\sqrt{2}}(|L\rangle+|R\rangle)( \cos\alpha|W\rangle + \sin\alpha|P\rangle),
	\end{eqnarray}
	with $|L\rangle$ and  $|R\rangle$ denoting the right and the left arms, respectively. To prepare such a pre-selected state, in Fig.~\ref{fig2},
	the initial state $|\psi_{\rm in}\rangle$ is put into the wave-particle toolbox. The action of the toolbox is to convert it to the state $|\psi_{\rm out} \rangle$.
	We then send it towards a 50:50 beam splitter (i.e. BS1) and this will produce the desired pre-selected state $|\Psi_{\rm i} \rangle$.

	Secondly, we choose the post-selected state as
	\begin{eqnarray}\label{pos-selestate}
	|\Psi_{\rm f}\rangle=\frac{1}{\sqrt2}(|L\rangle|W\rangle + |R\rangle|P\rangle).
	\end{eqnarray}
	Essentially, we want to perform a measurement that gives answer ``yes" whenever the state is $|\Psi_{\rm f}\rangle$, and answer ``no" when the state is
	orthogonal to $|\Psi_{\rm f}\rangle$. We consider only the cases where answer ``yes" is obtained. Such a measurement setup can be realized by the optical setup
	as in shown Fig.~\ref{fig2}. The post-selection consists of a beam splitter BS2
followed by the $\sigma^{1234}$ operator on the right arm, with
	\begin{equation}
	\sigma^{1234}=
	\begin{bmatrix}
	0&1&0&0\\
	1&0&0&0\\
	0&0&0&1\\
	0&0&1&0
	\end{bmatrix}.
	\end{equation}

On the right arm, the particle state can be converted to the wave state after the actions of BS2 and $\sigma^{1234}$, i.e.,
	\begin{eqnarray}
	|R\rangle|P\rangle &\xrightarrow{BS}& |R\rangle e^\frac{i\phi_1}{2}(\cos\frac{\phi_1}{2}|2\rangle-i\sin\frac{\phi_1}{2}|4\rangle)\nonumber\\
	&\xrightarrow{\sigma^{1234}}& |R\rangle e^\frac{i\phi_1}{2}(\cos\frac{\phi_1}{2}|1\rangle-i\sin\frac{\phi_1}{2}|3\rangle)\nonumber\\
	& \longrightarrow &|R\rangle|W\rangle.  \label{part_to_wave}
	\end{eqnarray}
We can now verify the result of our post-selection setup. By substituting
Eq. (\ref{part_to_wave}) into the post-selected state, we have
	\begin{eqnarray}
	|\Psi_{\rm f}\rangle = \frac{1}{\sqrt{2}}(|{L}\rangle|{W}\rangle+|{R}\rangle|{P}\rangle)\nonumber\\
\xrightarrow{BS,\sigma^{1234}}
	|\Psi_{\rm f_1}\rangle =\frac{1}{\sqrt{2}}(|L\rangle+|R\rangle|)W\rangle.
	\end{eqnarray}
The beam splitter BS3 is chosen as $|L\rangle \xrightarrow{BS3}  \frac{1}{\sqrt{2}}(|R\rangle-|L\rangle)$, $|R\rangle \xrightarrow{BS3}\frac{1}{\sqrt{2}}(|R\rangle+|L\rangle) \label{BS3}$,
 such that the state $|\Psi_{\rm f_1}\rangle$ turns to $|\Psi_{\rm f_2}\rangle=|R\rangle|W\rangle$, and the detector D2 does not click.
Finally, the action of the operator $X= |W\rangle\langle W|$ is such that only the wave state $|W\rangle$ is transmitted and the particle state $|P\rangle$ is reflected. Hence the detector D3 does not click, and the detector D1 clicks with certainty if the post-selected state is indeed $|\Psi_{\rm f}\rangle$.

This is the explicit calculation to show that with the proposed setup, the detector D1 always clicks if $|\Psi_{\rm f}\rangle$ is the post-selected state.
If any other state is chosen, there will be finite probability that the detectors D2 and D3 click. A photon starting in any state orthogonal to $|\Psi_f\rangle$ will either end
up at detector D2 or D3 and certainly will not fire D1.
So we can conclude that we have been able to design the setup, by introducing certain operators such that only the particular post-selected state gives D1 click
with 100\% probability. With this measurement setup the state $|\Psi_{\rm f}\rangle$ will certainly end up in the detector D1, and any state orthogonal to $|\Psi_{\rm f}\rangle$
will end up in the detectors D2 or D3. We only focus on the cases when the detector D1 clicks.

As we have known that in the context of pre- and post-selections the measurement strategy used is the \textit{weak measurement}  we try to perform suitable weak measurements
and extract information about the wave and particle aspects of the photon through these weak values.
	Following the quantum Cheshire cat proposal~\cite{rab2017entanglement}, which has allowed one to separate the properties of a particle from the particle itself,
	here we shall separate the wave and particle attributes of a quantum entity.
	We now move on to define various operators  which measure whether the wave and particle attributes are present in the left and right arms.	
	Explicitly, we have the operators
	\begin{eqnarray}\label{operators-P}
	&&\Pi^R_P=\Pi^R \otimes \Pi_P=|R\rangle\langle R|\otimes |P\rangle\langle P|, \nonumber\\
	&&\Pi^L_P=\Pi^L \otimes \Pi_P=|L\rangle\langle L|\otimes |P\rangle\langle P|, \label{operators_def}
	\end{eqnarray}
	which determines if particle attribute are there in the right and left arms, respectively. Similarly, the operators
	\begin{eqnarray}\label{operators-W}
	&&\Pi^R_W=\Pi^R \otimes \Pi_W=|R\rangle\langle R|\otimes |W\rangle\langle W|, \nonumber\\
	&&\Pi^L_W=\Pi^L \otimes \Pi_W=|L\rangle\langle L|\otimes |W\rangle\langle W|,
	\end{eqnarray}
	determines if the wave attribute are there in the right and left arms, respectively.
	
	Now the weak value of any observable $\hat{A}$ is given by
	\begin{eqnarray}\label{weak}
	\langle \hat{A}\rangle_w =\frac{ \langle \Psi_{\rm f}| \hat{A} |\Psi_{\rm i} \rangle}{ \langle\Psi_{\rm f}|\Psi_{\rm i} \rangle},
	\end{eqnarray}
	where  $|\Psi_{\rm i} \rangle$ and $|\Psi_{\rm f}\rangle$ are the pre-selected  and the post-selected states,  respectively.
We find that the weak values of these observables  in our setup are as follows:
	\begin{eqnarray}\label{weak-1}
	&&\langle \Pi^L_P\rangle_w =0, \;\;\;\langle\Pi^R_P\rangle_w =\frac{\sin\alpha}{\cos\alpha + \sin\alpha}, \nonumber\\
    &&\langle \Pi^R_W\rangle_w =0, \;\;\;
	\langle\Pi^L_W\rangle_w =\frac{\cos\alpha}{\cos\alpha +\sin\alpha}. \label{weak-val}
	\end{eqnarray}
	It may be emphasized that a non-vanishing weak value of a projector indicates whether the system has been in the particular state represented by that projector
	between the pre- and post-selections. Similarly, if the weak value of the projector is null, then the system has not been in that state between the pre- and
	post-selections.
	Based on the above result, we see that the particle property is zero in the left arm, whereas the wave property is zero in the right arm. Therefore,
	we can safely conclude that the particle property of the photon is constrained to the right arm, and the wave property of the photon is constrained to the left
	arm in such a pre- and post-selected setup. This indicates that the wave and particle properties of the single photon has been indeed spatially separated. Thus,
	with the help of suitable pre- and post-selections we can dismantle the wave and particle nature of a single photon. For $\alpha = \frac{\pi}{4}$, we have equal
	superposition of the wave and the particle state in the pre-selection, i.e.,
	$|\Psi_{\rm i} \rangle = \frac{1}{\sqrt{2}}(|L\rangle+|R\rangle) \otimes  \frac{1}{\sqrt{2}} ( |W \rangle + |P \rangle)$
 and the weak values for the particle and wave attributes are given by
	$\langle\Pi^R_P\rangle_w =\frac{1}{2}$ and $\langle\Pi^L_W\rangle_w =\frac{1}{2}$. In this case, half of the particle attribute is present in the right arm and half of the wave attribute is present in the left arm of the interferometer.

\emph{Complementarity.}---The above scheme is also applicable for
any quantum entity such as an electron or a neutron. One interesting point is that even though
we have been able to dismantle the wave and particle
properties of a single photon, this is actually consistent
with the complementarity principle that we will discuss here.  The quantum entity respects ``unity in diversity''.
Note that $\{ |L\rangle ,|R\rangle  \} \in \mathcal{H}^2$
 with $\Pi_{\mathcal{H}^2}=\Pi_L + \Pi_R = \openone$ (here $\openone$ being the $2\times 2$ unit matrix), and the modes in the wave-particle tool box $\{ |1\rangle, |2\rangle,  |3\rangle,  |4\rangle  \} \in \mathcal{H}^4$ with $\sum_{i=1}^4 \Pi_i =\sum_{i=1}^4 |i\rangle\langle i| = \openone\otimes \openone$. We can define another orthonormal basis $\{|W\rangle, |\Bar{W}\rangle, |P\rangle, |\Bar{P}\rangle \} \in  \mathcal{H}^4$ and with the resolution of identity as given by $\Pi_{\mathcal{H}^4}=\Pi_W + \Pi_{\Bar{W}}+ \Pi_P + \Pi_{\Bar{P}} =  \openone\otimes \openone$.
 With the pre-and post-selected states as given in (2) and (3), we have $\langle \Pi_{\mathcal{H}^2}\otimes \Pi_{\mathcal{H}^4}\rangle=1$, i.e.,
 \begin{eqnarray}
 \langle\Pi^L_W\rangle_w + \langle \Pi^L_P\rangle_w +\langle\Pi^L_{\Bar{P}}\rangle_w + \langle\Pi^L_{\Bar{W}}\rangle_w \nonumber\\
 +  \langle \Pi^R_W\rangle_w + \langle \Pi^R_{\Bar{W}}\rangle_w
 + \langle\Pi^R_P\rangle_w + \langle \Pi^R_{\Bar{P}}\rangle_w = 1. \label{complement2}
 \end{eqnarray}
 Further, we note that the weak values for various projectors satisfy these conditions:
\begin{eqnarray}
 \langle  \Pi^L_P\rangle_w =  \langle  \Pi^L_{\Bar{P}}\rangle_w = \langle \Pi^R_W\rangle_w = \langle \Pi^R_{\Bar{W}}\rangle_w =0, \nonumber \\
\langle\Pi^R_P\rangle_w =\frac{\sin\alpha}{\cos\alpha + \sin\alpha}, \;\;\; \langle \Pi^R_{\Bar{P}}\rangle_w =0, \nonumber\\
 \langle\Pi^L_W\rangle_w =\frac{\cos\alpha}{\cos\alpha +\sin\alpha},\;\;\  \langle \Pi^L_{\Bar{W}}\rangle_w =0.
 \end{eqnarray}
Therefore, we have
 \begin{eqnarray}
 \langle\Pi^R_P\rangle_w + \langle\Pi^L_W\rangle_w = 1.
 \end{eqnarray}
 This is a new complementarity relation between the wave and particle attributes in the weak measurement setting, i.e., the sum of the
 wave attribute in the left path and the particle attribute in the right path cannot be arbitrarily large.
 Interestingly, even though the wave and particle attributes have been dismantled, the prediction is consistent with the complementarity principle.

\section{Conclusion and Discussion}
The wave-particle duality is a fundamental concept of quantum mechanics, which implies that a physical entity is both a wave and a particle.
There has been a lot of debates regarding the wave-particle duality from the past, and it has been an interesting topic of research as well as one of the least understood aspects in quantum mechanics. Although this duality has worked well in physics to produce experimental confirmations, its interpretation is still under scanner. Though Niels Bohr have viewed such a duality as one aspect of the concept of the complementarity principle, there may be more to it. Here, in this work by exploiting the advantages of weak measurement and a pre- and post-measurement setup, we have spatially separated the so-called wave and particle attributes of a quantum entity. Even though they are dismantled, they still respect a new complementarity
relation. This also brings about some further fundamental questions like: What is this the wave aspect in the left arm of the interferometer like? How is it different from the general wave properties exhibited by an entity? Similarly we may also ask what is the ``solely particle'' aspect like in the right arm like?  It would be interesting to expect that, the interference fringes on the screen would vanish when one adopts the lights with solely particle property to perform the Young-type double-slit experiments, and also the electron diffraction effects would disappear when one adopts the electrons with solely particle property to perform the corresponding experiments.

In our work, we have realized the possibility of completely separating wave property and particle property for a quantum object. The proposal in our work is related to the quantum Cheshire cat, for which some physical attributes can be separated from the particle itself. In the next stage, we would like to furthermore consider a tripartite separation, i.e., the separation among the quantum object itself, the wave attribute, and the particle attribute. Once such a separation is achieved, then one will obtain a quantum Cheshire ``supercat''. We anticipate further experimental progresses in this direction in the near future.

	\begin{acknowledgments}
		J.L.C. is supported by National Natural Science Foundation of China (Grant Nos. 11875167, 12075001).
	\end{acknowledgments}
	
	\textbf{Disclosures}. The authors declare no conflicts of interest.

\textbf{Data availability}. No data were generated or analyzed in the presented research.

\end{document}